\documentclass[twocolumn,floatfix,superscriptaddress]{revtex4-1}
\usepackage{dcolumn}
\usepackage{graphicx}
\usepackage{amsmath,amssymb}
\usepackage{amsthm}
\usepackage{float}
\usepackage{placeins}
\usepackage{physics}
\usepackage{color}
\usepackage[colorlinks=true,allcolors=blue]{hyperref}
\begin{document}
\title{Aspects of holographic Langevin diffusion in the presence of anisotropic magnetic field}
\date{\today  \hspace{1ex}}
\author{Qi Zhou}
\email{qizhou@mails.ccnu.edu.cn}
\affiliation{Key Laboratory of Quark \& Lepton Physics (MOE) and Institute of Particle Physics,Central China Normal University, Wuhan 430079, China}
\affiliation{Helmholtz Research Academy Hessen for FAIR (HFHF),GSI
Helmholtz Center for Heavy Ion Physics. Campus Frankfurt, 60438
Frankfurt, Germany}
\affiliation{Institut f\"ur Theoretische Physik, Johann Wolfgang Goethe-Universit\"at,Max-von-Laue-Str.\ 1, D-60438 Frankfurt am Main, Germany}

\author{Ben-Wei Zhang}
\email{bwzhang@mail.ccnu.edu.cn}
\affiliation{Key Laboratory of Quark \& Lepton Physics (MOE) and Institute of Particle Physics,Central China Normal University, Wuhan 430079, China}

\begin{abstract}
We study the holographic Langevin diffusion coefficients of a heavy quark, when traveling through a strongly coupled anisotropic plasma in the presence of magnetic field~$\mathcal{B}$. 
While previous studies have characterized Langevin diffusion coefficients in the magnetic branes model, our analysis uncovers several new features among the five coefficients in this magnetic anisotropic plasma, advancing the understanding of heavy quark dynamics in such environments.
In particular, we clarified how motion velocities shape momentum broadening and its directional dependence.
It is observed that the transverse Langevin diffusion coefficients depend more on the direction of motion rather than the directions of momentum diffusion at the ultra-fast limit, while an opposite conclusion is found when the moving speed is sufficiently low.
For the Longitudinal Langevin diffusion coefficient, we find that motion perpendicular to~$\mathcal{B}$ affects the Langevin coefficients stronger at any fixed velocity.
We should also emphasize that all five Langevin coefficients are becoming larger with increasing velocity.
We find that the universal relation~$\kappa^{\parallel}>\kappa^{\perp}$ in the isotropic background, is broken in a different new case that a quark moving paralleled to~$\mathcal{B}$. 
This is one more particular example where the violation of the universal relation occurs for the anisotropic background.
Further, we find the critical velocity of the violation will become larger with increasing~$\mathcal{B}$.
\end{abstract}

\maketitle
\section{Introduction}
The heavy-ion collisions (HIC) experiments at the Relativistic Heavy Ion Collider (RHIC) and the Large Hadron Collider (LHC) are believed to create almost the most perfect fluid Quark Gluon Plasma (QGP)~\cite{BRAHMS:2004adc,PHENIX:2004vcz,PHOBOS:2004zne,STAR:2005gfr}. 
This provides a novel window for studying the physics of Quantum Chromodynamics (QCD) at a strongly coupled regime.
Since the properties of a strongly coupled system cannot be reliably calculated directly by perturbative techniques, one has to resort to some nonperturbative approaches to overcome the challenges.

The AdS/CFT correspondence~\cite{Maldacena:1997re,Witten:1998qj,Gubser:1998bc} is one of the very promising approaches to deal with these problems in QCD at strong couple scenario where can't be perturbatively handled properly~\cite{Witten:1998zw,Aharony:1999ti}.
On gravity side, the heavy quark diffusion in a strong coupled plasma can be understood as the fluctuation correlations of the trailing string. 
The studying of the stochastic nature of a heavy quark in a holography was proposed by~\cite{Gubser:2006nz,Casalderrey-Solana:2007ahi}. 
Soon the stochastic motion is formulated as a Langevin process~\cite{deBoer:2008gu,Son:2009vu}. 
Since the heavy quarks in HIC experiments are relativistic in many cases, the relativistic Langevin evolution are studied in~\cite{Giecold:2009cg,Casalderrey-Solana:2009ifi}
as well as in non-conformal frameworks in~\cite{Gursoy:2010aa,Kiritsis:2011bw,Kiritsis:2013iba} towards the multiple scales of QCD.
Recently, the contribution of nonlinear terms has been considered in \cite{Chakrabarty:2019aeu,Chakrabarty:2020ohe,Bu:2021jlp,Bu:2022oty} using a new prescription for Schwinger-Keldysh correlators proposed in \cite{glorioso2018prescription}.
The effect of the magnetic field directly on the heavy quark moving was studied in~\cite{Matsuo:2006ws,Kiritsis:2011ha}, which ignores the effect of the magnetic field on the plasma. 
It is studied that the heavy quark moving in magnetized plasma with~$\gamma M\leq \sqrt{\mathcal{B}}$~($\gamma$ is Lorentz factor and~$M$ is heavy quark mass), at strong magnetic field limit $e\mathcal{B}\gg T^2$~\cite{Mamo:2016xco,Li:2016bbh}. 
Other aspects of the influence of the magnetic field on heavy quarks using holography~\cite{Zhu:2021vkj,Zhang:2018pyr,Zhang:2018mqt,Zhang:2019jfq,Zhu:2019ujc,Zhu:2019igg,Zhou:2020ssi,Chen:2017lsf,He:2013qq,Dudal:2018rki,Giataganas:2018rbq,Iwasaki:2021nrz,Chakrabortty:2013kra,Rougemont:2020had,Fischler:2014tka,Rougemont:2014efa,Pourhassan:2017gza,Chakrabortty:2013kra,Fischler:2012ff,Tong:2012nf} and references therein.

An universal inequality relation, $\kappa_{\parallel}>\kappa_{\perp}$, has been found in the generic isotropic background with AdS/CFT correspondence~\cite{Giataganas:2013hwa}, based on the developments of the theory chosen in \cite{Gursoy:2010aa}. 
A way to violate this relation is to consider the Brownian motion in anisotropic background. 
And the fact that the transverse Langevin diffusion coefficients (LGV-coefficient) may be larger than a specific longitudinal LGV-coefficient, is found in both Janik-Witaszczyk \cite{Janik:2008tc} and Mateos-Trancanelli \cite{Mateos:2011ix,Mateos:2011tv} spatial anisotropic background in~\cite{Giataganas:2013zaa}. 
Due to the rapid expansion along the beam direction, the QGP experiences an anisotropic phase both in momentum and coordinate space before the system reaches the isotropic phase in a short period of time.
And the strong magnetic field is another important source of anisotropy, and it is interesting to study its influence on heavy quark diffusion at HIC.
To be specific we want to investigate how the anisotropy induced by an uniform magnetic field affects the both longitudinal and transverse LGV-coefficients. 

A strong magnetic field is produced in HIC experiments at RHIC around~$e\mathcal{B} \sim 0.02GeV^2$ 
and LHC up to~$e\mathcal{B} \sim 0.3 GeV^2$~\cite{Skokov:2009qp,Deng:2012pc,Bloczynski:2012en,Tuchin:2013ie,Bali:2011qj}, which has attracted much attention to study the influences on properties of QGP from the magnetic field. 
It's also known the existence of even stronger magnetic fields in the interior of the dense neutron stars~\cite{Duncan:1992hi} and early stages of the Universe~\cite{Vachaspati:1991nm,Grasso:2000wj}.
In weakly coupled quantum chromodynamics~(QCD) theory, the influences from the strong magnetic field~$\mathcal{B}$ to dynamic heavy quark has been studied at the Lowest-Landau-Level approximation, which for a thermal medium suggests the regime~$e\mathcal{B}\gg T^2$~\cite{Fukushima:2015wck,Kurian:2019nna,Sadofyev:2015tmb,Zhang:2020efz,Bandyopadhyay:2021zlm}.
The top-down magnetic branes model is dual to a deformed strong coupled SYM theory with a constant magnetic field~\cite{DHoker:2009mmn,DHoker:2012rlj,DHoker:2010onp}. 
Except for the top-down approach to modeling QCD for example \cite{DHoker:2009mmn,DHoker:2012rlj,DHoker:2010onp,Witten:1998xy,Hata:2007mb}, the bottom-up method is also very popular in this area and has yielded many extraordinary results~\cite{Erlich:2005qh,DaRold:2005mxj,Karch:2006pv,Gubser:2008ny,Gubser:2008yx,Gursoy:2008za,Gursoy:2010fj,Critelli:2017oub,Cai:2012xh,Giataganas:2017koz}. 
Although these models may not be exactly the same as the real QCD, it is expected that this kind of effort could help to reveal some intrinsic features of the plasma.

In this work, we choose to study LGV-coefficients within the widely studied magnetic branes model~\cite{DHoker:2009mmn,DHoker:2012rlj,DHoker:2010onp} to present our findings. 
The authors of \cite{Finazzo:2016mhm} first computed LGV-coefficients for the magnetic branes model and discovered several interesting properties of these coefficients.
In addition to confirming some conventional conclusions previously reported in \cite{Finazzo:2016mhm}, we conduct an in-depth investigation of heavy quark diffusion using LGV-coefficients in the presence of a uniform magnetic field, employing the same strategies.
We also aim to study heavy quark diffusion in an anisotropic background induced by a uniform magnetic field~$\mathcal{B}$, and the relationships between LGV-coefficients in an anisotropic background are another important motivation for this work.
Furthermore, we hope that our study can provide insights into transport simulations of hard probes in HIC.

This paper is organized as follows. 
In the next section section~\ref{sec2}, we will review the magnetic branes model and also the main procedures to deduce LGV-coefficients within the membrane paradigm.
We discuss all five LGV-coefficients and study the influences to relativistic LGV-coefficients from a uniform magnetic field at the different setups in section~\ref{sec3}. 
In section~\ref{sec4}, we give parameterized expressions to all five LGV-coefficients in the case of a strong magnetic field limit~$B\gg T$. 
The last part~\ref{sec5} is devoted to a short summary.

\section{The setups}\label{sec2}

The top-down magnetic branes model is dual to a deformed strong coupled SYM theory with a constant magnetic field.
The bulk action is a uniform Maxwell field coupled to 5 dimension Einstein gravity with a negative cosmological constant~\cite{DHoker:2009mmn},
\begin{equation}
S=\frac{1}{16\pi G_5}\int
d^5x\sqrt{-g}(R-\frac{12}{L^2}-F^2)+S_{body},
\label{action}
\end{equation}
where $G_5$ is the 5 dimension gravitational constant and $F$ stands for the Maxwell field strength 2-form, the radius of the $AdS_5$ spacetime $L$ is set to unity for the rest of the paper. $S_{body}$ contains the Chern-Simons terms, Gibbons-Hawking terms and other contributions necessary for a well posed variational principle. There is no sense of confinement for the nonexistence of dilaton field.

Following the general ansatz in~\cite{DHoker:2009mmn}, the anisotropic magnetic branes background with a uniform magnetic field is given as
\begin{equation}
ds^2=-U(r)dt^2+e^{2V(r)}(dx^2+dy^2)+e^{2W(r)}dz^2+\frac{dr^2}{U(r)}
\label{metric}
\end{equation}
for metric. And for field strength~$F$, one has
\begin{equation}
F=Bdx\wedge dy,
\end{equation}
with constant $B$ for the magnetic field strength along axis-$z$ direction. The horizon is located at $r\sim r_h$ and the boundary is located at $r\rightarrow\infty$. The three functions $U(r)$, $V(r)$, and $W(r)$ are obtained by solving the equations of motion.

Maxwell’s equations following from (\ref{action}) are automatically satisfied by the ansatz (\ref{metric}), while the set of linearly independent components read
\begin{equation}
U(V^{\prime\prime}-W^{\prime\prime})+(U^\prime+U(2V^\prime+W^\prime))(V^\prime-W^\prime)=-2B^2e^{-4V}\label{e1},
\end{equation}
\begin{equation}
2V^{\prime\prime}+W^{\prime\prime}+2(V^\prime)^2+(W^\prime)^2=0,\label{e2}
\end{equation}
\begin{equation}
\frac{1}{2}U^{\prime\prime}+\frac{1}{2}U^{\prime}(2V^\prime+W^\prime)=4+\frac{2}{3}B^2e^{-4V},\label{e3}
\end{equation}
\begin{equation}
2U^\prime V^\prime+U^\prime W^\prime+2U(V^\prime)^2+4UV^\prime
W^\prime=12-2B^2e^{-4V}.
\label{e4}
\end{equation}

A charged system will undergo a dimensional reduction in the presence of strong fields due to the projection towards the lowest Landau level~\cite{Shovkovy:2012zn,Gusynin:1995nb}.
The magnetic branes solution must satisfy two sets of boundary conditions, which corresponds to a holographic renormalization group flow interpolating between a~$BTZ\times R^2$ near horizon solution and a near boundary~$AdS_5$ asymptotic solution at $r\rightarrow\infty$.
On the one hand, the geometry (\ref{metric}) in IR should reduce to a~$BTZ$ black hole~\cite{Banados:1992wn} times a two dimensional torus~$T^2$ in the spatial directions transverse to the magnetic field.
The exact analytical solution near the horizon~($r\sim r_h$) was found as
\begin{equation}
ds^2=-\frac{r^2f(r)}{\mathcal{R}^2}dt^2+\mathcal{R}^2B(dx^2+dy^2)+\frac{r^2}{\mathcal{R}^2}dz^2+\frac{\mathcal{R}^2}{r^2f(r)}dr^2,
\end{equation}
with $f(r)=1-\frac{r_h^2}{r^2}$ and $\mathcal{R}=\frac{L}{\sqrt{3}}$ refers to the the $BTZ$ black hole radius. 
On the other hand, there should be a near boundary $AdS_5$ asymptotic solution at $\tilde{r}\rightarrow\infty$,
\begin{equation}
    ds^2=r^2(-dt^2+dx^2+dy^2+dz^2)+\frac{dr^2}{r^2},
    \label{metric-ultaviolate}
\end{equation}
for one we must recover the dynamics of $\mathcal{N} = 4$ SYM without the influence of the magnetic field at UV.

Following the procedure mentioned in \cite{DHoker:2009mmn},
here one constructs a numerical solution that interpolates between the near horizon $BTZ\times T^2$ and asymptotic~$AdS_5$ in the boundary. 
To fix the horizon at $\tilde{r}_h=1$, one use rescaled coordinates $\tilde{r}$ so that also give
\begin{equation}
\tilde{U}(1)=0.
\end{equation}
With a rescaled $\tilde{t}$, one take 
\begin{equation}
\tilde{U}^{\prime}(1)=1,
\end{equation}
which sets the temperature to a fixed value $T=\frac{1}{4\pi}$ and leave $B$ as the free parameter.
Further one take 
\begin{equation}
   \tilde{V}(1)=\tilde{W}(1)=0 
\end{equation} 
with the help of rescaled $\tilde{x}$, $\tilde{y}$, $\tilde{z}$. 

With these conditions, (\ref{e1}) and (\ref{e3}) give us the initial data
\begin{equation}
\tilde{V}^\prime(1)=4-\frac{4b^2}{3},\qquad
\tilde{W}^\prime(1)=4+\frac{2b^2}{3}.
\label{vpwp}
\end{equation}
where $b$ stands for the value of the magnetic field in the rescaled coordinates. 

\begin{figure}
\centering
\includegraphics[width=8.8cm]{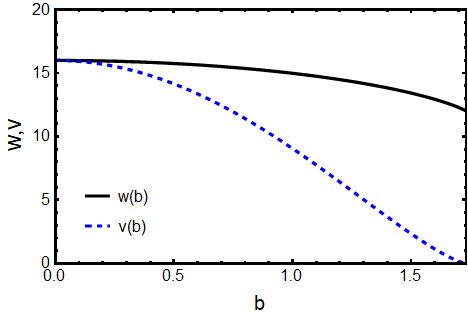}
\caption{$v(b)$~(dash curve) and $w(b)$~(solid curve) against $b$.}
\label{figure:wv}
\end{figure}

Integrating the ODE out from $\tilde{r}=1$ to a large value of $\tilde{r}$, one find geometry will have the asymptotic behavior at
$\tilde{r}\rightarrow \infty$,
\begin{equation}
\tilde{U}(\tilde{r})\rightarrow \tilde{r}^2, \qquad e^{2\tilde{V}(\tilde{r})}\rightarrow v(b) \tilde{r}^2, \qquad e^{2\tilde{W}(\tilde{r})}\rightarrow
w(b) \tilde{r}^2,
\end{equation}
where  $v(b)$ and $w(b)$ are parameters which can be determined numerically. We plot two key coefficients~$w(b)$ and~$v(b)$ in Fig.(\ref{figure:wv}) as a function of~$b$, which is match to Fig.(1) in~\cite{Critelli:2014kra}.

Since the solution should have an asymptotic $AdS_5$ at the UV, one can introduce coordinate
\begin{equation}
(x,y,z)\rightarrow
(\tilde{x}/\sqrt{v(b)},\tilde{y}/\sqrt{v(b)},\tilde{z}/\sqrt{w(b)},
\end{equation}
then he full bulk metric~(\ref{metric}) after these rescaling reads
\begin{equation}
\begin{split}
\label{metric-scaled}
ds^2=&-\tilde{U}(\tilde{r})d\tilde{t}^2\\
&+\frac{e^{2\tilde{V}(\tilde{r})}}{v(b)}(d\tilde{x}^2+d\tilde{y}^2)
+\frac{e^{2\tilde{W}(\tilde{r})}}{w(b)}d\tilde{z}^2
+\frac{d\tilde{r}^2}{\tilde{U}(\tilde{r})}\\
F=&\frac{b}{v(b)}d\tilde{x}\wedge d\tilde{y}.
\end{split}
\end{equation}

One can obtain the thermodynamics of the gauge theory from (\ref{metric-scaled}).
And the rescaled magnetic field~$b$ is related to the bulk magnetic field~$B$ by
\begin{equation}
B=\frac{b}{v(b)}.
\end{equation}
Since the geometry has to be asymptotic $AdS_5$ at UV, the first equation in (\ref{vpwp}) imply~$0\leq b\leq\sqrt{3}$.
The physically observable magnetic field at the boundary~$\mathcal{B}=\sqrt{3}B$, as argued in~\cite{DHoker:2009mmn} by comparing the Chern-Simons term in (\ref{metric}) with the $\mathcal{N} = 4$
SYM chiral anomaly. By using the fact that any physical quantity in this model should depend on the dimensionless quantity, ratio~$T/\sqrt{\mathcal{B}}$, one fixs the temperature at~$T=1/(4\pi)$ thus compute
\begin{equation}
    \frac{T}{\sqrt{\mathcal{B}}}=\frac{3^{-\frac{1}{4}}}{4\pi}\sqrt{\frac{v(b)}{b}}.
\end{equation}

Next, we compute the LGV-coefficients of a heavy quark in a uniform magnetic field induced anisotropy medium mainly by following the author of \cite{Giataganas:2013hwa} and also \cite{Finazzo:2016mhm,Gubser:2006nz,Gursoy:2010aa}.
We consider a general diagonal metric in radial coordinates,
\begin{equation}
    ds^2=g_{tt}dt^2+g_{xx}dx^2+g_{yy}dy^2+g_{zz}dz^2+g_{rr}dr^2,
\end{equation}
where the boundary is~$r\rightarrow\infty$. For simplicity, one set~$g_{xx}=g_{yy}$ similar to magnetic branes geometry (\ref{metric}).
Here, we consider a quark moving at direction~$x_p|_{(x_p=x,z)}$ and transverse to direction~$x_k$, and calculate the LGV-coefficients transverse to~$\mathcal{B}$ direction and along~$\mathcal{B}$ direction respectively. 

Holographically, the moving heavy quark of infinite mass on the boundary CFT correspond to the endpoint of the trailing string. The string dynamics are captured by the Nambu-Goto action
\begin{equation}
\begin{split}
    \label{nbaction}
     S_{NG}=-\frac{1}{2\pi\alpha'}\int \mathrm{d}\sigma^{\alpha} \mathrm{d}\sigma^{\beta} \sqrt{-\mathrm{det} g_{\alpha\beta}}, \\
     g_{\alpha\beta}=g_{\mu\nu}\frac{\partial X^{\mu}}{\partial \sigma^{\alpha}}\frac{\partial X^{\nu}}{\partial \sigma^{\beta}}.
\end{split}
\end{equation}
where $g_{\alpha\beta}$ is the induced metric, and $g_{\mu\nu}$ and $X_{\mu}$ are the branes metric and target space coordinates.

A moving heavy quark with a constant velocity $v$ has the usual parametrization
\begin{equation}
    \label{gauge}
    t=\sigma^{\alpha}, \quad r=\sigma^{\beta},\quad x=vt+\xi(r),
\end{equation}
where $\xi$ is the profile of the string in the bulk.
Then it's is deduced the world-sheet metric
\begin{equation}
\label{diag}
    g_{\alpha\beta}=
\left(
        \begin{matrix}
            g_{tt}+v^2g_{pp} &  g_{pp}v\xi' \\
             g_{pp}v\xi' & g_{rr}+g_{pp}\xi^{'2} 
        \end{matrix}
\right)
\end{equation}
One can find the critical point $r_c$ from $g_{\alpha\alpha}(r_c) = 0$, which means we find $r_c$ by solve the
\begin{equation}
    g_{tt}=-v^2g_{pp}.
\end{equation}
$C\equiv \frac{\partial \mathcal{L}}{\partial \xi}$ proportional to radial conjugate momentum.

One find the effective temperature $T_{ws}$ of the world-sheet horizon by diagonal world-sheet metric.
\begin{equation}
\label{tran_to_ws}
    d\sigma^{\alpha}\rightarrow d\sigma^{\alpha}+\frac{g_{\alpha\beta}}{g_{\alpha\alpha}}d\sigma^{\beta}.
\end{equation}
Further $h_{\alpha\beta}$ is the diagonal induced world-sheet metric given
\begin{equation}
    \begin{aligned}
            h_{\beta\beta} = \frac{g_{tt}g_{pp}g_{rr}}{g_{tt}g_{pp}+C^2}\\ h_{\alpha\alpha}=g_{tt}+g_{pp}v^2.
    \end{aligned}
\end{equation}
As derived by the authors of \cite{Giataganas:2013hwa,Giataganas:2013zaa}, the effective temperature reads,
\begin{equation}
\begin{aligned}
    T^{2}_{ws}&=\frac{1}{16\pi^2}\abs{\frac{g^{'2}_{tt}-v^4g^{'2}_{pp}}{g_{tt}g_{pp}}}\bigg|_{r=r_{c}}\\
    &=\frac{1}{16\pi^2}\abs{\frac{1}{g_{tt}g_{rr}}(g_{tt}g_{pp})'(\frac{g_{tt}}{g_{pp}})'}\bigg|_{r=r_{c}}.
    \label{tws}
\end{aligned}
\end{equation}

Considering the fluctuation in classical trailing string, one has 
\begin{equation}
    t=\sigma^{\alpha}, r=\sigma^{\beta}, x_{p}=vt+\xi(r)+\delta x_p(\sigma^{\alpha},\sigma^{\beta}),
\end{equation}
where the fluctuation of the form $\delta x_p(\sigma^{\alpha},\sigma^{\beta})$ along longitudinal and transverse direction of $x_p|_{(1,2,3)}$.
The action capture fluctuations around the trailing string reads
\begin{multline}
\label{2eaction}
    S_{f}=-\frac{1}{2\pi\alpha'}\int d\tau d\sigma \frac{H^{\alpha\beta}}{2}\\
    \times(N[r]\partial_{\alpha}\delta x_p\partial_{\beta}\delta x_p+\sum g_{ii}\partial_{\alpha}\delta x_i\partial_{\beta}\delta x_i),\\
    N(r)\equiv\frac{g_{tt}g_{pp}+C^2}{g_{tt}+g_{pp}v^2},\\
    H^{\alpha\beta}=\sqrt{-det(h)}h^{\alpha\beta}.\\
\end{multline}
Taking advantage of the membrane paradigm~\cite{Iqbal:2008by}, as derived by the authors of \cite{Giataganas:2013hwa,Giataganas:2013zaa}, one can directly reads LGV-coefficients without solving motion equation,
\begin{equation}
\label{kappat}
\begin{split}
\kappa_{\perp} &= \frac{g_{kk}}{\pi \alpha'}\bigg|_{r=r_c} T_{ws},
\end{split}
\end{equation}
and
\begin{equation}
\label{kappal}
\begin{split}
\kappa_{\parallel} &= \frac{1}{\pi\alpha'}\frac{( g_{tt}g_{pp})'}{g_{pp} (\frac{g_{tt}}{g_{pp}})'}\bigg|_{r=r_c} T_{ws}.
\end{split}
\end{equation}
We rederived the formulas for our Langevin diffusion coefficients Eq. (\ref{kappat}), Eq. (\ref{kappal}), as well as the crucial intermediate formulas Eq. (\ref{tran_to_ws}) and Eq. (\ref{tws}), by following \cite{Giataganas:2013hwa,Giataganas:2013zaa} where the authors explained that these formulas are clear and explicit. 

In the case of magnetic black branes, one can directly reads from the metric (\ref{metric-scaled}) as
\begin{align}
    &g_{tt}=-U(r)=-\tilde{U}(\tilde{r})\label{gtt},\\
    &g_{xx}=g_{yy}=e^{2V(r)}=\frac{e^{2\tilde{V}(\tilde{r})}}{v(b)}\label{gxx},\\
    &g_{zz}=e^{2W(r)}=\frac{e^{2\tilde{W}(\tilde{r})}}{w(b)}\label{gzz},\\
    &g_{rr}=\frac{1}{U(r)}=\frac{1}{\tilde{U}(\tilde{r})}\label{grr}.
\end{align}

When the quark moving along~$\mathcal{B}$, one have a longitudinal LGV-coefficients $\kappa^{v\parallel B}_{\parallel}$ and one transverse LGV-coefficient $\kappa^{v\parallel B}_{\perp}$. 
Inserting (\ref{gzz}) into (\ref{tws}), one have effective temperature~$T^\parallel_{ws}$
\begin{equation}
\begin{aligned}
    T^\parallel_{ws}
    &=\frac{1}{4\pi}\left[\abs{\left(\frac{e^{2\tilde{W}(\tilde{r})}\tilde{U}(\tilde{r})}{w(b)}\right)'\left(\frac{w(b)\tilde{U}(\tilde{r})}{e^{2\tilde{W}(\tilde{r})}}\right)'}\right]^{\frac{1}{2}}\bigg|_{r=r_c}.
\end{aligned}
\end{equation}
With~$g_{pp}=g_{zz}$ and~$g_{kk}=g_{xx}$, one the expressions into (\ref{kappal}) and (\ref{kappat}), then get 
\begin{equation}
\begin{split}
    \kappa^{v\parallel B}_{\parallel}&=\frac{T^{\parallel}_{ws}w(b)}{\pi\alpha'e^{2\tilde{W}(\tilde{r})}}
    \left(\frac{e^{2\tilde{W}(\tilde{r})}\tilde{U}(\tilde{r})}{-w(b)}\right)'
    \left(\frac{-w(b)\tilde{U}(\tilde{r})}{e^{2\tilde{W}(\tilde{r})}}\right)^{'-1}\bigg|_{r=r_c},\\
    \kappa^{v\parallel B}_{\perp}&=\frac{T^{\parallel}_{ws}}{\pi\alpha'}\frac{e^{2\tilde{W}(\tilde{r})}}{w(b)}\bigg|_{r=r_c}.
    \label{lgv-para}
\end{split}
\end{equation}

When the quark moving transverse to~$\mathcal{B}$, one also have a longitudinal LGV-coefficients $\kappa^{v\perp B}_{\parallel}$ but two transverse LGV-coefficient $\kappa^{v\perp B}_{(\perp,\parallel)}$ and~$\kappa^{v\perp B}_{(\perp,\perp)}$, for the anisotropy in plane transverse to motion. We denote $\kappa^{v\perp B}_{(\perp,\parallel)}$ as the transverse LGV-coefficient when quark moving perpendicular to the direction of~$\mathcal{B}$ and diffusion paralleled to the direction of motion. We also denote $\kappa^{v\perp B}_{(\perp,\perp)}$ as the transverse LGV-coefficient when moving and diffusion direction perpendicular to the direction of~$\mathcal{B}$.
Inserting (\ref{gxx}) into (\ref{tws}), we have effective temperature
\begin{equation}
\begin{aligned}
    T^\perp_{ws}&=
    &=\frac{1}{4\pi}\left[\abs{\left(\frac{e^{2\tilde{V}(\tilde{r})}\tilde{U}(\tilde{r})}{v(b)}\right)'\left(\frac{v(b)\tilde{U}(\tilde{r})}{e^{2\tilde{V}(\tilde{r})}}\right)'}\right]^{\frac{1}{2}}\bigg|_{r=r_c}.
\end{aligned}
\end{equation}
With~$g_{pp}=g_{xx}$ and~$g_{kk}=g_{zz}$, one inserts the expressions into (\ref{kappal}) and (\ref{kappat}), then get  
\begin{equation}
\begin{split}
    \kappa^{v\perp B}_{\parallel}&=\frac{T^{\perp}_{ws}v(b)}{\pi\alpha'e^{2\tilde{V}(\tilde{r})}}
    \left(\frac{e^{2\tilde{V}(\tilde{r})}\tilde{U}(\tilde{r})}{-v(b)}\right)'
    \left(\frac{-v(b)\tilde{U}(\tilde{r})}{e^{2\tilde{V}(\tilde{r})}}\right)^{'-1}\bigg|_{r=r_c},\\
    \kappa^{v\perp B}_{(\perp,\parallel)}&=\frac{T^{\perp}_{ws}}{\pi\alpha'}\frac{e^{2\tilde{V}(\tilde{r})}}{v(b)}\bigg|_{r=r_c}.
    \label{lgv-perppara}
\end{split}
\end{equation}
With~$g_{pp}=g_{xx}$ and~$g_{kk}=g_{yy}$, one the inserts expressions into (\ref{kappat}), then get
\begin{equation}
    \kappa^{v\perp B}_{(\perp,\perp)}=\frac{T^{\perp}_{ws}}{\pi\alpha'}\frac{e^{2\tilde{W}(\tilde{r})}}{w(b)}\bigg|_{r=r_c}.
    \label{lgv-perpperp}
\end{equation}

In isotropic conformal limit, one can obtain the well-known results~\cite{Gubser:2006nz,Casalderrey-Solana:2007ahi}
\begin{equation}
\begin{split}
    \kappa_{\perp}^{SYM}=\sqrt{\lambda} \pi T^3 \gamma^{\frac{1}{2}},\\
    \kappa_{\parallel}^{SYM}=\sqrt{\lambda} \pi T^3 \gamma^{\frac{5}{2}}.
    \label{lgvsym}
\end{split}
\end{equation}
The (\ref{lgv-para}), (\ref{lgv-perppara}) and (\ref{lgv-perpperp}) reduce to this isotropic results after~$\mathcal{B}$ is turned off.

\section{Numeric Results}\label{sec3}
\subsection{Directions dependence}\label{section:direction}
\begin{figure}[!htbp]
    \includegraphics[width=8.6cm]{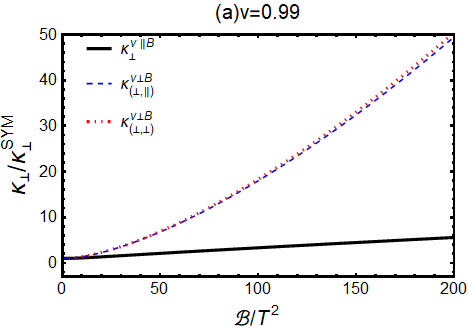}
    \includegraphics[width=8.6cm]{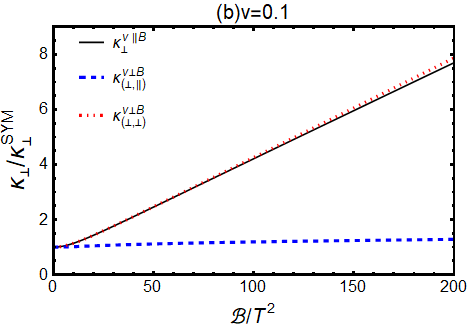}
    \caption{The transverse LGV-coefficients~$\kappa^{v\parallel B}_{\perp}$,~$\kappa^{v\perp B}_{(\parallel,\perp)}$ and~$\kappa^{v\perp B}_{(\perp,\perp)}$, at~$v=0.99$~(Upper panel) and at~$v=0.1$~(Lower panel), as a function of $\mathcal{B}/T^2$ are normalized by the conformal limit.}
    \label{figure:ktvsb}
\end{figure}

\begin{figure}[!h]
    \includegraphics[width=8.6cm]{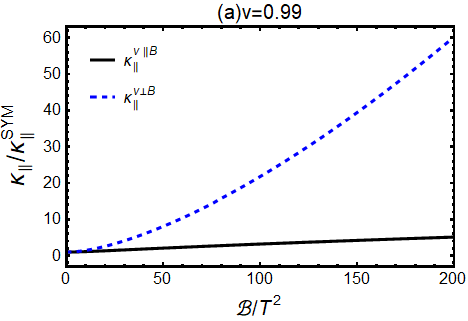}
    \includegraphics[width=8.6cm]{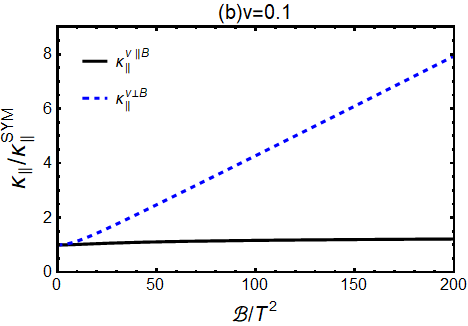}
    \caption{The longitudinal LGV-coefficients~$\kappa^{v\parallel B}_{\parallel}$ and~$\kappa^{v\perp B}_{\parallel}$, at~$v=0.99$~(Upper panel) and at~$v=0.1$~(Lower panel), as a function of $\mathcal{B}/T^2$ are normalized by the conformal limit.}
    \label{figure:klvsb}
\end{figure}

Initially, in this subsection, we will examine two extreme cases, in the ultra fast limit and at a fixed sufficiently small speed, to study the relations between different LGV-coefficients in all scale with~$\mathcal{B}/T^2$, which is the only dimensionless parameter in this model.
The numerical results of transverse LGV-coefficients~$\kappa_{\perp}$ and longitudinal LGV-coefficients~$\kappa_{\parallel}$ as a function of magnetic field~$B/T^2$ are displayed in Fig.~\ref{figure:ktvsb} and Fig.~\ref{figure:klvsb}, normalized by the conformal limit given in (\ref{lgvsym}).
Fig.~\ref{figure:ktvsb} shows transverse LGV-coefficients~$\kappa^{v\parallel B}_{\perp}$, $\kappa^{v\perp B}_{(\perp,\parallel)}$ and~$\kappa^{v\perp B}_{(\perp,\perp)}$, while Fig.~\ref{figure:klvsb} show longitudinal LGV-coefficients~$\kappa^{v\parallel B}_{\parallel}$ and $\kappa^{v\perp B}_{\parallel}$.
When moving at the ultra fast limit, as shown in plot~(a) in Fig.~\ref{figure:ktvsb}, the transverse LGV-coefficients of quark $\kappa_{\perp}$ has a relation reads $\kappa_{\perp}^{v\parallel B} \leq \kappa^{v\perp B}_{(\perp,\parallel)} \leq \kappa^{v\perp B}_{(\perp,\perp)}$ and $\kappa^{v\perp B}_{(\perp,\parallel)} \sim \kappa^{v\perp B}_{(\perp,\perp)}$, which means the transverse LGV-coefficients of quark mostly depending on the direction of moving but not the direction of transverse momentum diffusion when the moving velocity approaching to light speed. 
Since $\kappa^{v\perp B}_{(\perp,\parallel)}$ and $\kappa^{v\perp B}_{(\perp,\perp)}$ share the same motion direction, perpendicular to the magnetic field.
It also means that the anisotropy induced by a uniform magnetic field in plane~$xoz$ contributes little to the strength of transverse LGV-coefficients when moving in the ultra-fast limit.
But when the moving velocity is sufficiently small, $v=0.1$ is taken for example, one will find an opposite conclusion that the transverse LGV-coefficients of quark depend more on the direction of transverse momentum diffusion rather than the direction of moving. Since one can find $\kappa^{v\perp B}_{(\perp,\perp)} \sim \kappa^{v\parallel B}_{(\perp)}$ on plot (b) in Fig.(\ref{figure:ktvsb}) and they share the same transverse momentum diffusion direction perpendicular to the magnetic field (We will further examine this conclusion in the next subsection).  

As for the longitudinal LGV-coefficients shown in Fig.~\ref{figure:klvsb}, we consistently find that $\kappa_{\parallel}^{v\perp B} >\kappa_{\parallel}^{v\parallel B}$ both in ultra fast limit and at a fixed sufficiently small speed, which means moving perpendicular to $\mathcal{B}$ direction affects the longitudinal LGV-coefficients more strongly compared to motion along $\mathcal{B}$ direction. 
By comparing transverse LGV-coefficients in Fig.~\ref{figure:ktvsb} with the longitudinal LGV-coefficients in Fig.~\ref{figure:klvsb}, we find that the transverse LGV-coefficients are more sensitive to the moving velocity in the anisotropic magnetic background than the longitudinal LGV-coefficients.

\subsection{Velocity dependence}
\begin{figure}[!htbp]
    \includegraphics[width=8.8cm]{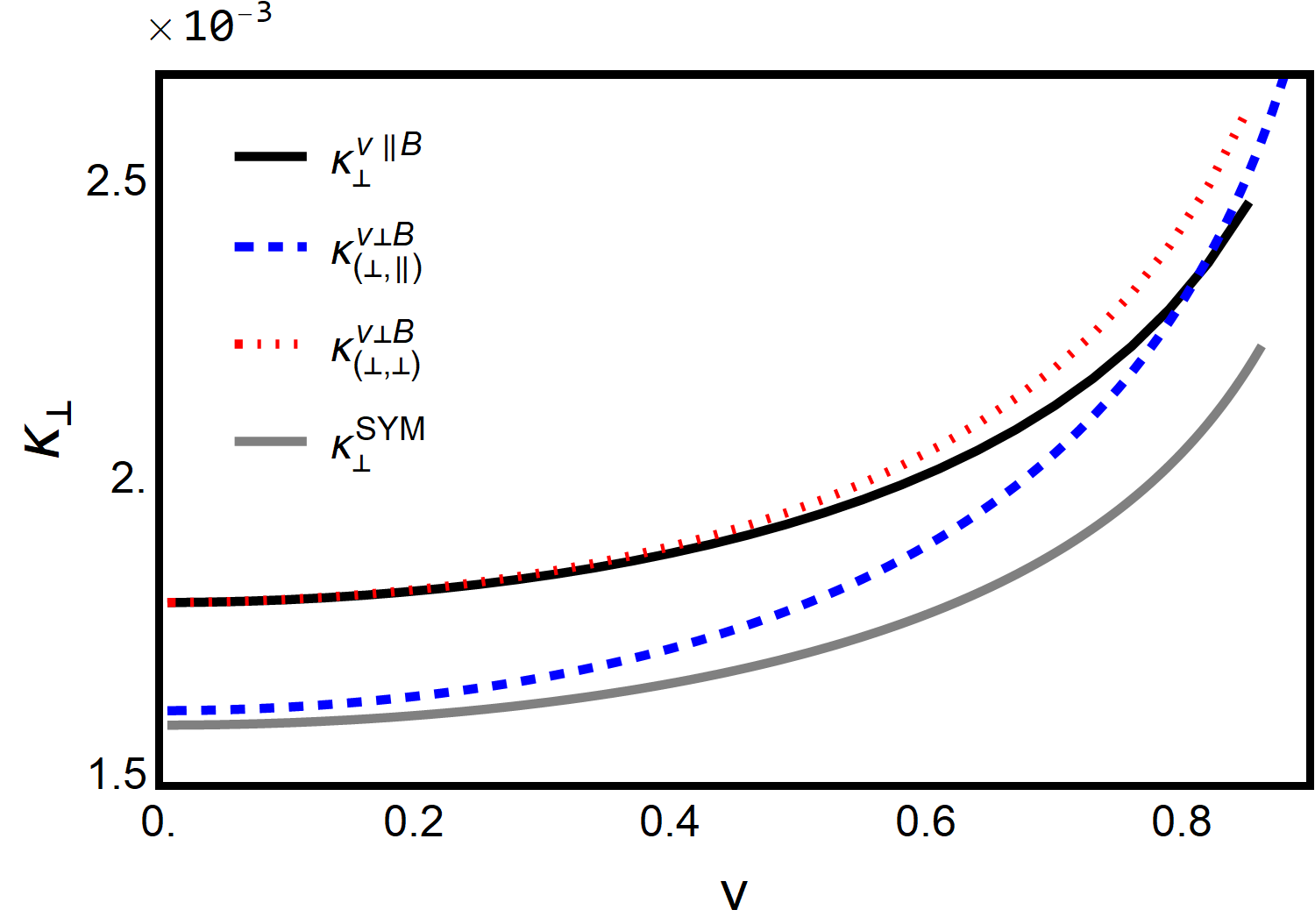}
    \includegraphics[width=8.8cm]{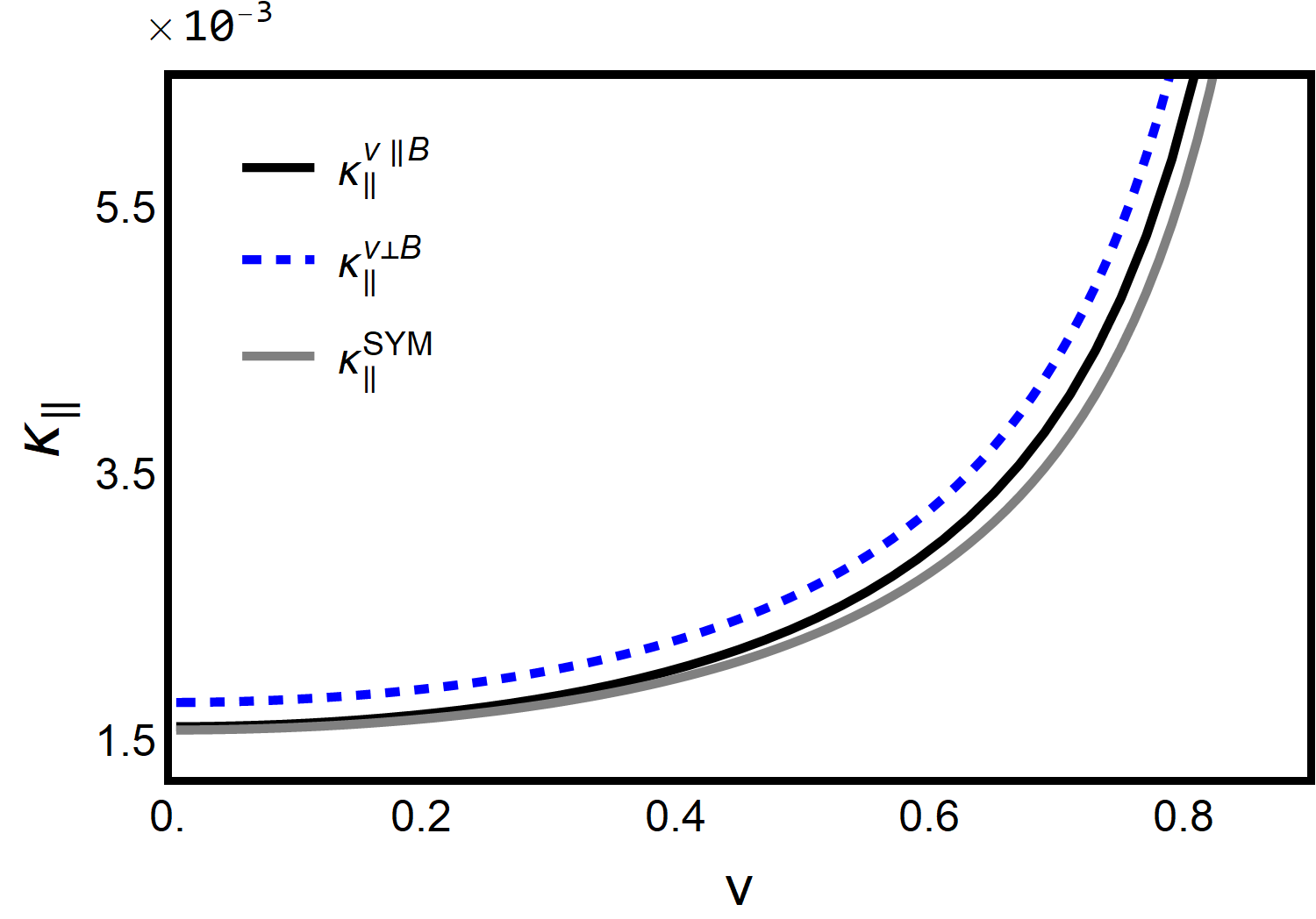}
    \caption{The transverse LGV-coefficients $\kappa_{\perp}$ are shown in the upper panel, while the longitudinal LGV-coefficients $\kappa_{\parallel}$ are shown in the lower panel. The corresponding conformal limit result is also included as a function of the motion velocity $v$.}
    \label{figure:kvsv}
\end{figure}

We further examine the impact of moving velocity $v$ on both transverse and longitudinal LGV-coefficients. In Fig.~\ref{figure:kvsv}, we plot the LGV-coefficients as a function of quark velocity~$v$, accompanied by the corresponding results in the conformal limit given in (\ref{lgvsym}).
It demonstrates that the increasing moving velocity~$v$ enhances the LGV-coefficients in all cases we mentioned in this anisotropic magnetic background, which is very similar to the results in the conformal limit. 
This conclusion is also similar to the outcomes in spatially anisotropic case in~\cite{Giataganas:2013zaa}. 

Unlike the two extreme cases with fixed speeds in subsection~\ref{section:direction}, we can study the same question with varying moving speeds.
It is found that, at a sufficiently small moving speed, the transverse momentum broadening of a heavy quark depends more on the direction of momentum diffusion rather than the direction of moving with both $\kappa^{v\perp B}_{(\perp,\parallel)} < \kappa^{v\parallel B}_{\perp} \leq \kappa^{v\perp B}_{(\perp,\perp)}$ and $ \kappa^{v\perp B}_{(\perp,\perp)}\sim \kappa^{v\parallel B}_{\perp}$ in the upper panel of Fig.~(\ref{figure:kvsv}). 
($\kappa^{v\perp B}_{(\perp,\perp)}$ and $\kappa^{v\parallel B}_{\perp}$ share the same transverse diffusion direction, perpendicular to the magnetic field.)
It is also found that the longitudinal LGV-coefficients, shown in the lower panel of Fig.~\ref{figure:kvsv}, $\kappa_{\parallel}^{v\perp B} >\kappa_{\parallel}^{v\parallel B}$ at all fixed speeds.

\subsection{Violations of the longitudinal to transverse LGV-coefficient ratios}
\begin{figure}[!htbp]
    \includegraphics[width=8.6cm]{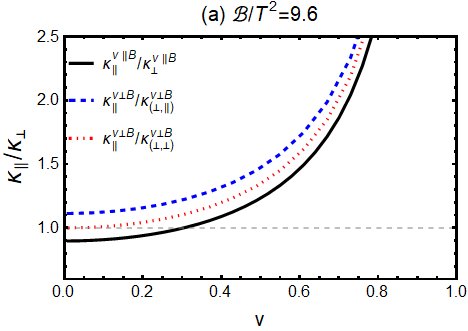}
    \includegraphics[width=8.8cm]{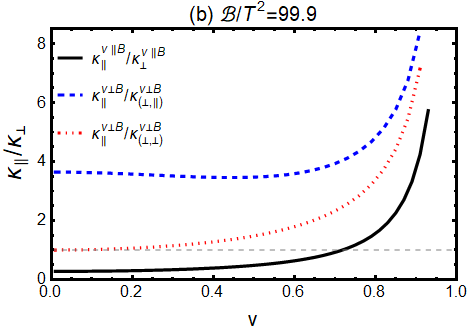}
    \caption{The ratios of~$\kappa_{\parallel}/\kappa_{\perp}$ as a function of quark motion velocity~$v$ in different case. The only violation is~$\kappa^{v\perp B}_{\parallel}/\kappa^{v\perp B}_{(\perp,\perp)}$ at both~$\mathcal{B}/T^2=9.6$~(Upper panel) and~$\mathcal{B}/T^2=99.9$~(Lower panel).}
    \label{figure:ratio}
\end{figure}

With both transverse and longitudinal LGV-coefficients, the universal equation~\cite{Gursoy:2010aa} $\kappa_{\parallel}>\kappa_{\perp}$ and its violation~\cite{Giataganas:2013hwa} can be examined within this specific case of a top-down magnetic branes model.
Fig.~\ref{figure:ratio} presents numerical results of the ratios of longitudinal over transverse LGV-coefficients, $\kappa^{\parallel}/\kappa^{\perp}$, as a function of quark moving velocity $v$.
One can find that~$\kappa_{\parallel} < \kappa_{\perp}$ at sufficiently low speed, which violates the universal equation $\kappa_{\parallel}>\kappa_{\perp}$ found in isotropic background.
In our study, the violation occurs at sufficiently low speeds only when moving parallel to the direction of a magnetic field. 
This represents another instance of violating the universal relation.
Since \cite{Giataganas:2013zaa} reports violated fractions for $\kappa^{v \perp B}_{\parallel} / \kappa^{v \perp B}_{(\perp, \parallel)}$ (using consistent notation) in a prolate anisotropic background, this differ from our case where the violation occurs for $\kappa^{v \parallel B}_{\parallel} / \kappa^{v \parallel B}_{\perp}$.
In conclusion, the violation of the universal relation $\kappa_{\parallel} > \kappa_{\perp}$ in any anisotropic background, as predicted in \cite{Giataganas:2013hwa}, also holds in this specific magnetic brane model.
However, which specific components 
$\kappa^{v \parallel B}_{\parallel} / \kappa^{v \perp B}_{(\perp, \parallel)}$, 
$\kappa^{v \perp B}_{\parallel} / \kappa^{v \perp B}_{(\perp, \parallel)}$, or 
$\kappa^{v \perp B}_{\parallel} / \kappa^{v \perp B}_{(\perp, \perp)}$ are violated depends on the metric's details, and identifying which component is violated requires further understanding of the quantitative details of anisotropy in metric.


By comparing plot~(a) and plot~(b) in Fig.~\ref{figure:ratio}, one can further take an investigation the influences on the violation from increasing magnetic field strength~$\mathcal{B}$. 
We find the critical velocity $v$ of violation become larger with increasing magnetic strength~$\mathcal{B}$, then we can conclude that the relationship, $\kappa_{\parallel}>\kappa_{\perp}$, will eventually be violated as long as the magnetic field is strong enough in this magnetic anisotropic plasma, as in spatial anisotropic case~\cite{Giataganas:2013zaa}.
LGV-Coefficients are critical quantities in describing heavy quark transport in QGP \cite{Moore:2004tg,Uphoff:2012gb}, contributing to the study of heavy flavor jet quenching in HIC (see \cite{Xu:2017obm,Li:2021xbd,Li:2024uzk} and references therein). 
We hope the studies of violations between LGV-Coefficients could shed new light on these simulations.

\section{Near horizons analytical limit}\label{sec4}

One has a special geometry in the case of the strong magnetic field $B\gg T$, which is a product of 3 dimension $BTZ$ and trivial flat 2 dimension as mentioned in section~\ref{sec2}. 
The exact solution near the horizon ($r\sim r_h$), which denotes the product of a~$BTZ$ black hole times a two dimensional torus $T^2$, was found as
\begin{equation}
ds^2=-\frac{r^2f(r)}{\mathcal{R}^2}dt^2+\mathcal{R}^2B(dx^2+dy^2)+\frac{r^2}{\mathcal{R}^2}dz^2+\frac{\mathcal{R}^2}{r^2f(r)}dr^2,
\end{equation}
with $f(r)=1-\frac{r_h^2}{r^2}$ and $\mathcal{R}=\frac{L}{\sqrt{3}}=\frac{1}{\sqrt{3}}$ referring to the the $BTZ$ black hole radius. The horizon is located at $r = r_h$  and the boundary located at~$r \rightarrow\infty$. The Hawking temperature~$T$ of the $BTZ$ black hole~$T=\frac{3r_h}{2\pi}$.

One has
\begin{equation}
\begin{split}
    g_{tt}(r)&=-3(r^2-r_h^2),\\
    g_{xx}(r)&=g_{yy}(r)=\frac{B}{3}, \\
    g_{zz}(r)&=3r^2,\\
    g_{rr}(r)&=\frac{1}{3(r^2-r_h^2)}.
    \label{btzfraction}
\end{split}
\end{equation}

By repeating the procedure in section~\ref{sec2}, one has critical point~$r^{\parallel}_c$ when a heavy quark moving paralleled to magnetic field $B$, as 
\begin{equation}
    r^{\parallel}_c=\frac{r_h}{\sqrt{1-v^2}}.
\end{equation}

The effective temperature of 2 dimensional black hole of a quark feel denoted as~$T_{s}^{\parallel}$ when quark moving along and transverse to magnetic field~$B$,
\begin{equation}
    T^{\parallel}_{s}=T\sqrt{1+v^2}.
\end{equation}

Asserting (\ref{btzfraction}) to (\ref{tws}), (\ref{kappal}) and (\ref{kappat}), one get longitudinal and perpendicular LGV-coefficients as
\begin{equation}
    \kappa^{v\parallel B}_{\parallel}=\frac{2 \pi \sqrt{\lambda} T^3 \sqrt{v^2+1}}{3 (v-1)^2}+\frac{2 \pi \sqrt{\lambda} T^3 \sqrt{v^2+1}}{3 (v+1)^2}.
    \label{btzlgv1}
\end{equation}
and
\begin{equation}
    \kappa^{v\parallel B}_{\perp}=\frac{\sqrt{v^2+1}}{3 \pi }\sqrt{\lambda}B T
    \label{btzlgv2}
\end{equation}

Similar to the parallel case, computations of the LGV-coefficients for a quark moving perpendicular to a magnetic field~$\mathcal{B}$ can be done. The critical point is at 
\begin{equation}
    r^{\perp}_c=\frac{\left(B v^2+4 \pi ^2 T^2\right)^{\frac{1}{2}}}{3} .
\end{equation}
The effective temperature temperature reads
\begin{equation}
    T^{\perp}_{s}=\frac{\sqrt{B v^2+4 \pi ^2 T^2}}{2 \pi }.
\end{equation}

The perpendicular and longitudinal LGV-coefficients reads
\begin{equation}
\begin{split}
    \kappa^{v\perp B}_{\parallel}=\frac{B^{3/2} \sqrt{\lambda} v}{6 \pi ^2}+\frac{\sqrt{B} \sqrt{\lambda} T^2}{3 v}
    \label{btzlgv3}
\end{split}
\end{equation}
and
\begin{equation}
    \kappa^{v\perp B}_{(\perp,\parallel)}=\frac{B^{3/2} \sqrt{\lambda} v^3}{6 \pi ^2}+\sqrt{B} \sqrt{\lambda} T^2 v,
    \label{btzlgv4}
\end{equation}
and
\begin{equation}
    \kappa^{v\perp B}_{(\perp,\perp)}=\frac{B^{3/2} \sqrt{\lambda} v}{6 \pi ^2}+\frac{\sqrt{B} \sqrt{\lambda} T^2}{3 v}.
    \label{btzlgv5}
\end{equation}

It should be mentioned that the perpendicular LGV-coefficients (\ref{btzlgv2}), (\ref{btzlgv4}) and (\ref{btzlgv5}) have been calculated to compute jet quenching parameters~$\hat{q}$ in \cite{Li:2016bbh}. 
In present study, we recompute these quantities including the two additional longitudinal LGV-coefficients (\ref{btzlgv1}) and (\ref{btzlgv3}) for completeness. 
In the case of the strong magnetic fie $B\gg T$, the final results of (\ref{btzlgv3}), (\ref{btzlgv4}) and (\ref{btzlgv5}) have ignore the $o(T^2/B)$ in our calculation. We recommend \cite{Li:2016bbh} for discussion about $\kappa_{\perp}$. The longitudinal LGV-coefficients $\kappa^{v \parallel B}_{\parallel}$ is independent of $B$ and similar to weak coupling results of $\kappa_{\parallel}\propto \alpha^2_{s}T^3$ introduced in \cite{Fukushima:2015wck}. 
However, it's demonstrated from (\ref{btzlgv5}) that $\kappa^{v \perp B}_{\parallel}$ depends on $B$ as $\kappa \propto B^{3/2}$ at leading order. 

\section{Conclusion}\label{sec5}

The study of heavy quark transport properties, such as Langevin diffusion coefficients, as functions of parameters like temperature, chemical potential, momentum anisotropy, and magnetic field strength \(\mathcal{B}\), is crucial for characterizing and understanding the QGP.
In the present work, we take an investigation on relativistic heavy quark diffusion in strongly coupled anisotropic plasma with uniform magnetic field. 
The Langevin diffusion coefficients are calculated within the membrane paradigm in the magnetic branes model which has been extensively studied to investigate the magnetic effects on various observables in strongly coupled QCD scenarios.
We present several new dynamic properties of Langevin diffusion of a heavy quark in strongly coupled anisotropic plasma with uniform magnetic field in this work, advancing the understanding of heavy quark dynamics in such environments. In particular, we clarified how motion velocities shape momentum broadening and its directional dependence.
  
To be more specific, we find several new interesting features among the five Langevin diffusion coefficients in the magnetic anisotropic plasma, besides confirming the conclusions found in \cite{Finazzo:2016mhm}.   
By comparing LGV-coefficients for quark motion in different directions, we clarified how motion velocities shape momentum broadening and its directional dependence.
It is observed that the transverse LGV-coefficients depend more on the direction of motion than the direction of diffusion at the ultra-fast limit, while one would find an opposite conclusion when it comes to a lower moving speed.  
For longitudinal LGV-coefficients, we find that motion perpendicular to~$\mathcal{B}$ consistently affects the LGV-coefficients more strongly at any fixed velocity.
We find all the LGV-coefficients becoming larger with increasing speed, which is similar to the behaviors in the case of both pure SYM theory \cite{Gubser:2006nz,Casalderrey-Solana:2007ahi} and axion-driven anisotropy theory \cite{Giataganas:2013zaa}.
We also conclude that the transverse LGV-coefficients are more sensitive to the moving velocity in the anisotropic magnetic background than the longitudinal LGV-coefficients.

Our conclusions are consistent with \cite{Finazzo:2016mhm}, which was the first to consider all five LGV coefficients in the magnetic branes model. They found that LGV-coefficients are enhanced in the presence of an external magnetic field relative to the zero magnetic field case. They also found that momentum diffusion in directions transverse to the magnetic field is generally larger than in the direction of the field.  
Four of the LGV-coefficients, which were scaled in the zero magnetic field case, increase with increasing velocity, in contrast to the coefficient $\kappa^{v\perp B}_{\perp v}$ which behaves differently at large $v$. 
 
We confirm that the universal relation $\kappa_{\parallel} > \kappa_{\perp}$ is also violated in the magnetic brane model, as expected in \cite{Giataganas:2013hwa}. 
However, this violation differs from the pattern found in the spatial anisotropy case discussed in \cite{Giataganas:2013zaa}. 
We also find the critical velocity of the violation will become larger with increasing~$\mathcal{B}$ in this magnetic background as in spatial anisotropic case \cite{Giataganas:2013zaa}.
We should emphasize that, the anisotropic metric is the main contribution to the violation behaviors of Langevin diffusion coefficients not the sources themselves, as explained in \cite{Giataganas:2013hwa}. In a sense, any such metric (induced by magnetic fields or other sources) may affect Langevin coefficients similarly. 
Furthermore, our discussion, supported by a comparison with \cite{Giataganas:2013zaa}, indicates that identifying the specific violated component in anisotropic background requires understanding the metric's anisotropy details.

To provide a more complete analysis in the strong magnetic field limit $B \gg T^2$, we compute all five LGV coefficients in the BTZ background, building upon the work of \cite{Li:2016bbh}, which focused on the three transverse components. 

\section*{ACKNOWLEDGMENTS}
Q. Zhou would like to thank E. Bratkovskaya and the PHSD team for their warm hospitality during part of this work.
We would also like to thank D. Giataganas for pointing out a mistake.
This research is supported by the Guangdong Major Project of Basic and Applied Basic Research No. 2020B0301030008, and Natural Science Foundation of China with Project Nos. 11935007. 
This research was supported in part by National Natural Science Foundation of China~(NSFC) under Project No. 12535010 and by Guangdong Basic and Applied Basic Research Foundation under Project No.2022A1515110392.

\bibliography{ref}
\end{document}